\documentclass[aps,epsfig,twocolumn,superscriptaddress,showpacs,floatfix]{revtex4}
\usepackage[bf]{caption}
\usepackage{epsfig}
\usepackage{amsmath}
\usepackage{wrapfig}

\begin{document}

\title{Enhancement of kinetic energy fluctuations due to expansion}
\author{A. Chernomoretz}
\email{ariel@df.uba.ar}
\affiliation{ Departamento de F\'{i}sica, Universidad de Buenos Aires,\\
Buenos Aires, Argentina.}
\author{F. Gulminelli}
\altaffiliation{Member of the Institut Universitaire de France.}
\affiliation{LPC Caen (IN2P3-CNRS/Ensicaen 
et Universit\'e) F-14050 Caen C\'edex.
France}
\author{M.J. Ison}
\affiliation{ Departamento de F\'{i}sica, Universidad de Buenos Aires,\\
Buenos Aires, Argentina.}
\author{C.O. Dorso}
\affiliation{ Departamento de F\'{i}sica, Universidad de Buenos Aires,\\
Buenos Aires, Argentina.}

\date{\today}

\begin{abstract}
Global equilibrium fragmentation inside a freeze out constraining volume
is a working hypothesis widely used in nuclear fragmentation statistical models.
In the framework of classical Lennard Jones molecular dynamics,
we study how the relaxation of the 
fixed volume constraint affects the 
posterior evolution of microscopic correlations, and how a non-confined
fragmentation scenario is established.                                        
A study of the  dynamical evolution of the relative kinetic energy fluctuations
 was also performed. We found that asymptotic measurements of such  observable
can be related to the number of decaying channels available to the system at
fragmentation time.

\end{abstract}


\maketitle

\section{Introduction}\label{sintro}
The {\small multifragmentation} phenomenon has attracted the attention of the
intermediate/high energy nuclear community since more than a decade.
Starting in the 1980's, a lot of attention has been paid to the
identification and characterization of the occurrence of a nuclear
liquid-gas phase transition. 
On one hand, indications of critical behavior
were obtained assuming a Fisher-like scaling relation for fragment mass
distributions \cite{purdue,jbelliott,search_crit1,nuovo}. On the
other, it has been claimed that the observed critical behavior in static
observables is compatible with a first order phase transition
\cite{prlgulmi,pan,gulmichom,gulmi_fluc,cvdagostino}. 

As in nuclear fragmentation experiments one only has access to asymptotic 
observables, the fragmentation stage of the dynamic evolution must be 
reconstructed using simplified models.
To accomplish this task, the most widely used statistical models of the 
MMMS \cite{mmmc2} or SMM \cite{smm} type, 
replace the intractable problem of quantal N-body correlations with a 
non interacting gas of fragments confined with sharp boundary conditions
\cite{mmmc2,smm}.
 A distinctive feature of these equilibrium fragmentation models is that a 
vapor-branch can be recognized in the system caloric curve (CC). 
Moreover, in the statistical sampling of configurations, 
the observed asymptotic flux is exclusively associated with the mean 
Coulomb energy of partitions inside the constraining volume.
Collective motions are completely disregarded before and
during the fragment formation stage.
 
On the contrary, when fragmentation is considered in a non-constrained
free-to-expand scenario qualitatively different features are found. In
this case a local equilibrium picture replaces the global one, 
and a characteristic flattening of the caloric curve at high energy 
can be found~\cite{noneqfrag}. 
This happens as a direct consequence of the presence of a
collective degree of freedom: the expansive motion behaves like a
heat sink, and precludes the disordered kinetic energy associated with
the system temperature to increase without bounds~\cite{ale97}.
These last remarks rise some reasonable concerns regarding the sharp
constraining volume working hypothesis that are worth to be analyzed.

In recent contributions the study of the behavior of the
relative kinetic energy fluctuations, $A_{K}\equiv N\sigma _{K}^{2}/<K>^{2}$,
has received considerable attention. It has
been shown~\cite{gulmichom,chomazprl} that partial energy fluctuations can
be used to measure the EOS, and to extract information of the possible phase
transition occurring in finite systems like multifragmenting nuclei.
Negative specific heats are expected to signal first order phase transitions
in these systems~\cite{grossbook,cvdagostino}. Taking into consideration that
kinetic energy fluctuations and the system's specific heat are related by~\cite
{lebowitz}: 
\begin{equation}
{\frac{1 }{N}} <\sigma_K^2>_E={\frac{3}{{2 \beta^2}}} (1 - {\frac{3 }{{2 C}}%
})  \label{eqFluc}
\end{equation}
(where $\beta=1/kT$) negative values of the specific heat are expected to
appear whenever $\sigma_K^2$ gets larger than the corresponding canonical
expectation value: ${\frac{3 N}{{2 \beta^2}}}$.

Using a classical molecular dynamics model it has been already 
shown in~\cite{phasediagram} that
large relative kinetic fluctuations do take place in confined classical
isolated systems, provided that the constraining volumes are large enough
to accommodate non-overlapping configurational fragments. 
In this paper we want to pursue with the system's characterization by
studying its temporal evolution once we remove the constraining
walls within which it was equilibrated.
Our aim is to understand how much memory of the initial configuration
is kept in the asymptotic fragmentation patterns which are accessible
experimentally.
To that end we will focus on the dynamical evolution of fragment mass 
distribution functions (MDF), and the relative kinetic energy fluctuation,
$A_K$, that turns out to be a useful tool to probe the unconstrained 
fragmentation scenario.

\section{The model}\label{smodel}
The system under study is composed by excited drops made up of 147 particles
interacting via a 6-12 Lennard Jones potential with a cut-off radius
$r_{c}=3\sigma $. 
Energies are measured in units of the potential well ($\epsilon $), 
$\sigma $ characterizes the radius of a particle and $m$ is its mass.
We adopt adimensional units for energy,length and time such that $\epsilon=\sigma=1$, 
 $t_{0}=\sqrt{\sigma^{2}m/48\epsilon }$. 
Our particles are uncharged and the inclusion of Coulomb is certainly
a necessary step to insure that our results are pertinent to the nuclear
fragmentation problem~\cite{misoncoulomb}. However, 
the fact that the the nuclear force 
and the 
van der Waals interaction both share the same general features (i.e. short range
repulsion plus a longer range attraction) supports the use of this
simplified classical model to obtain qualitatively meaningful results in the
nuclear case, and serves also to explore phenomena of greater generality. 
The initial condition is chosen as a spherical confining `wall',
introducing an external potential $V_{wall}\sim (r-r_{wall})^{-12}$ with
a cut off distance $r_{cut}=1\sigma $. The set of classical equations of motion were
integrated using the well known velocity Verlet algorithm, which preserves
volume in phase space\cite{frenkel}.

Using this system, we have performed the following molecular
dynamics (MD) experiments. First, we let the system equilibrate inside the
constraining volume at a chosen density. We call this
initial condition the \emph{constrained
state}. Afterwards, we remove the constraining walls, and let the system
evolve for a very long time, such that the fragment composition reaches
stability. This will be referred to as the \emph{asymptotic stage}.
\\

One of the key observables to study phase transformations from a 
morphological point of view  is the fragment mass distribution. 
The simplest and more intuitive 
cluster definition is based on correlations in configuration space: a particle $i$ 
belongs to a cluster $C$ if there is another particle $j$ that belongs to $C$ and 
$|\mathbf{r_{i}}-\mathbf{r_{j}}|\leq r_{cl}$, where $r_{cl}$ is a parameter called the 
clusterization radius. In this work we took $r_{cl}=r_{cut}=3\sigma $. The algorithm 
that recognizes these clusters is known as the ``Minimum Spanning Tree'' (MST). 
The main drawback of this method is that only correlations in 
\textbf{r}-space are used.
MST clusters give no
meaningful information about the fragmentation dynamics when the system is
dense.

A more robust algorithm is based on the analysis of the so called ``Most
Bound Partition'' (MBP) of the system, introduced in Ref.~\cite{ecra}. The
MBP is the set of clusters $\{C_{i}\}$ for which the sum of the fragment
internal energies attains its minimum value: 
\begin{eqnarray}
{\{C_{i}\}} & = & \hbox{argmin}%
\scriptstyle {{\{C_{i}\}} } \textstyle {\left[E_{\{C_{i}\}}=%
\sum_{i}E_{int}^{C_{i}}\right]} \nonumber \\
E_{int}^{C_{i}} & = &\sum_{j\in C_{i}}K_{j}^{cm}+\sum_{{j,k\in C_{i}}\,{%
j\le k} }V_{jk}  \label{eq:eECRA}
\end{eqnarray}
where the first sum in (\ref{eq:eECRA}) is over the clusters of the
partition, $K_{j}^{cm}$ is the kinetic energy of particle $j$ measured in
the center of mass frame of the cluster which contains particle $j$, and $%
V_{ij}$ stands for the inter-particle potential. It can be shown that
clusters belonging to the MBP are related to the most-bound density
fluctuation in \textbf{r-p} space~\cite{ecra}. The algorithm that finds the
MBP is known as the ``Early Cluster Recognition Algorithm'' (ECRA). 

\begin{figure}[tbp]
\epsfig{figure=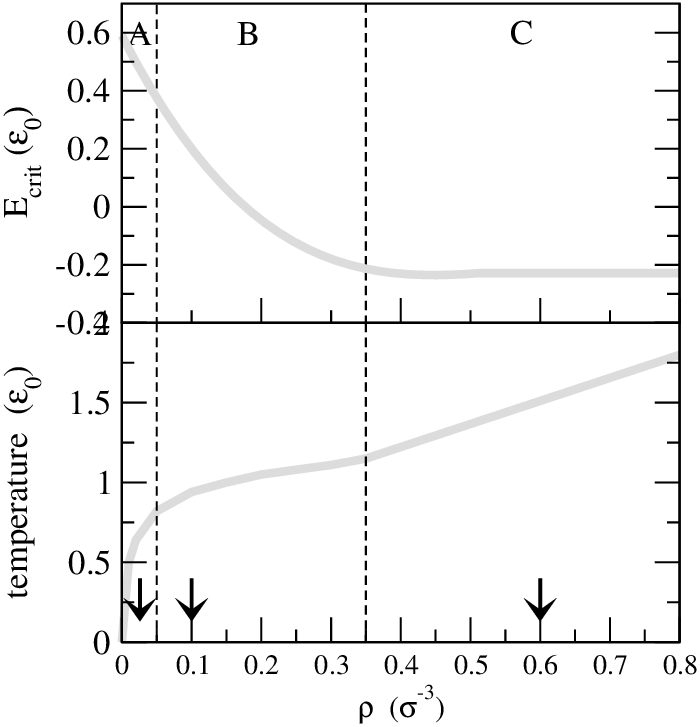,width=6cm}
\caption{Upper panel: energy $E_{crit}$ at which transition signals are
detected in a confined, finite size fragmenting system, as a function of the
system density. Lower panel: same transition line in the temperature-density
plane (see text for details). The arrows indicate the density values
considered for the present calculations.}
\label{figES}
\end{figure}

In a previous contribution we studied the EOS of a finite, constrained, and
classical system using cluster distributions properties in phase space and
in configurational space~\cite{phasediagram}. Figure~\ref{figES}
summarizes our findings. In the upper panel we show the density dependence of
the system energy $E_{crit}$ at which transition signals (power-law mass
spectrum, maxima in the second moment of the mass distribution, and
normalized mean variance) are detected. Given a confined system with a
density value $\rho $, for $E<E_{crit}(\rho )$ a big ECRA cluster can be
found, whereas for $E>E_{crit}(\rho )$ a regime with
high multiplicity of light ECRA clusters dominates. The
lower panel shows the same transition line in the most usual $(T,\rho )$
plane (see \cite{phasediagram} for details).

According to the calculated critical exponents $\tau $, and $\gamma $, three
density regions could be identified. One (very 
low density limit, region labeled 
$A$ in Figure~\ref{figES}) in which 
the boundary conditions are almost ineffective, and a backbending of the 
caloric curve can consequently be observed even in the isochore ensemble
\cite{imfm}. 
Another one (region $C$ in Figure~\ref{figES}),
corresponding to the high density regime, in which fragments in phase space
display critical behavior of 3D-Ising universality class type. And an
intermediate density region (region $B$ in Figure~\ref{figES}), in which
power-laws are displayed but can not be associated to the above mentioned
universality class (see Ref.~\cite{phasediagram}). 
The thermodynamic critical point of the liquid-gas phase transition
($\rho_c\approx .35\sigma ^{-3}$,$T_c\approx 1.12\epsilon_0$)  
corresponds to the meeting point of these two last regions.

Having these results in mind, we chose the following density values, $\rho
_{1}=0.026\sigma ^{-3}$, $\rho _{2}=0.10\sigma ^{-3}$, $\rho _{3}=0.6\sigma
^{-3}$ (i.e. one from each relevant region of the EOS) in order to study the
effect of the dynamic evolution on the observables that characterize the
initially equilibrated confined systems. \\

\section{Time Evolution of Particle correlations}
In order to explore the effects of constrain removal on the
fragment spectra of the systems under study, we compare the
confined ECRA-fragment mass distribution of the
constrained system, against the corresponding MST asymptotic ones.

\begin{figure}[tbph]
\epsfxsize=8cm \center{\leavevmode\epsfbox{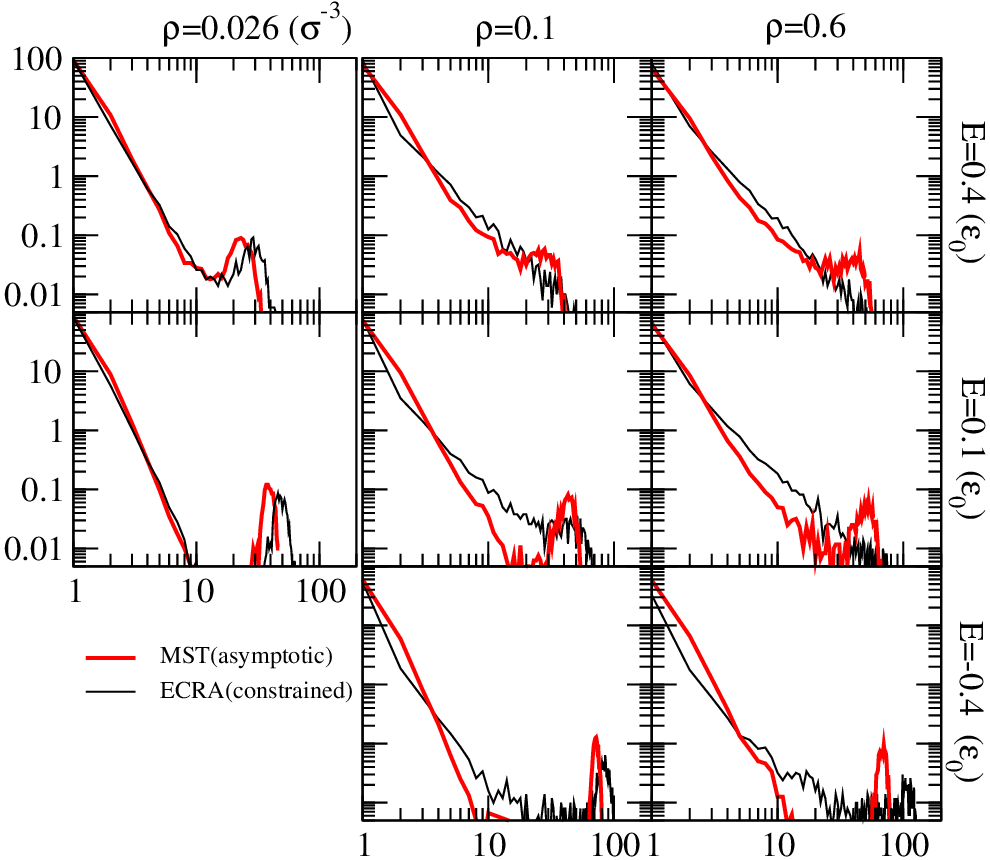}}
\caption{Confined ECRA vs asymptotic MST mass spectra calculated for 3
different initial densities: 
$\rho =0.026\sigma ^{-3},0.1\sigma ^{-3}$, and $0.6\sigma ^{-3}$ 
shown in first, second, and third columns respectively.} 
\label{figecramst}
\end{figure}

In Figure~\ref{figecramst} we show MDF calculations performed over systems
initially constrained at the density values above mentioned (from left to right: 
$\rho _{1},\rho _{2}$, and $\rho_{3}$) with different total energies (increasing $E$ 
from bottom to top).
For a given set of values $(E,\rho )$ the confined state ECRA-fragment mass 
spectra and the
corresponding asymptotic MST-fragment mass spectra can be compared.
It should be kept in mind that for $\rho_2$ and $\rho_3$ initial states
present just one big fragment according to the MST algorithm.

For the case of an initially extremely diluted system (first column in
Figure~\ref{figecramst}) confined states are already
fragmented in configurational space. The corresponding asymptotic fragment
structure remains essentially unaltered, aside of the evaporation of light
clusters from the heaviest fragment.

On the other hand, it can be seen that for initial higher densities ($\rho
_{2}$, and $\rho _{3}$) the $\mathbf{q}\!-\!\mathbf{p}$ correlations of the
confined state, probed by ECRA partitions, are modified by the dynamics that
follows the walls removal. For these densities, the size of the biggest
clusters detected in both, the confined and asymptotic stages, seem to agree
to a high degree. However noticeable differences are found in relative
abundances, specially for intermediate size fragments (IMF). 

The development of an expansive flux modifies the correlation
pattern observed in phase space for initially dense confined configurations
acting as a memory loosing mechanism of their initial fragment
structure. 
This result implies that even a subcritical density relatively 
diluted as the $\rho=0.1\sigma^{-3}$ case cannot be taken as a freeze out
configuration, since freeze out by definition implies a persistency of the
information up to asymptotic times.
It is then clear that between confined states and corresponding
asymptotic configurations, the intermediate stage, characterized by the
development of expansive collective motion
and the consequent expansion of the system, plays a crucial role in the
process of fragment formation. 
 The relative similarity between the asymptotic distributions originated
from the $\rho=0.1\sigma^{-3}$ and $\rho=0.6\sigma^{-3}$ initial conditions,
suggests that the expansion dynamics leads to a fragmentation pattern 
essentially determined by the total energy. On the other hand the distribution
originated from the initially most diluted system is sensibly different from
the denser cases. Indeed the size of the heaviest fragment is smaller and the
IMF distribution is steeper in this case. This means that a fragmentation
pattern originated from an expansion process is not equivalent to an
equilibrated freeze out with sharp boundary conditions. It has been
proposed\cite{flow} that such a fragmentation pattern may still be described in
statistical terms through information theory with the constraints of a
fluctuating volume and a radial flow. The possible pertinence of such a scenario
is left to future investigations.
\\

\section{Kinetic Energy fluctuations}
As was stated in the introduction, fragmenting finite systems at
microcanonical equilibrium display negative specific heats whenever relative
kinetic energy fluctuations surpass the canonical expected value. 
This is theoretically expected in the liquid-gas phase transition
of small systems if the volume is not constrained by sharp boundary
conditions\cite{houches}. 
Consistently it has been suggested that the magnitude $A_{K}$ can
be used to identify the occurrence of first order phase transitions in such
systems~\cite{gulmichom}.
\begin{figure*}[tbph]
\epsfxsize=14cm \center{\leavevmode\epsfbox{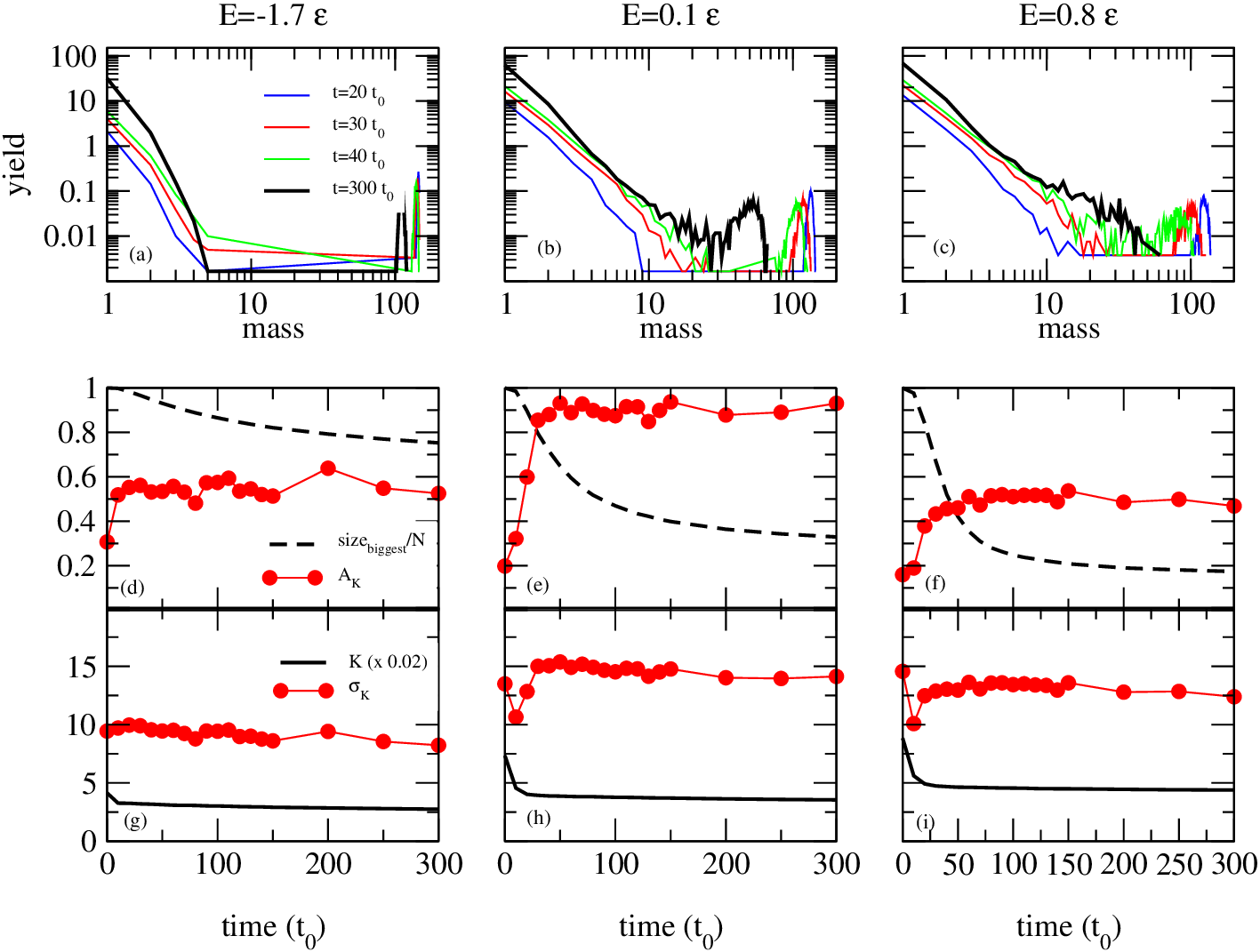}}
\caption{ Upper panels (3.a, 3.b, 3.c): 
MST fragment spectra calculated at $t=20,30,40$ and, 
$300t_{0}$ for a $\rho =0.6 \sigma^{-3}$ system at three different energies. 
Middle panels (3.d, 3.e, 3.f): $A_{K}(t)$ (red circles) and MST-biggest cluster 
mass (dashed lines) as a function of time. 
Lower panels (3.g, 3.h, 3.i): Total kinetic 
energy (solid line) and Kinetic energy fluctuations (red circles) 
as a function of time. }
\label{figakt}
\end{figure*}
If the equiprobability of the different configuration microstates
is assumed, as it is by construction the case in our initial conditions,
$K=K(T)$, and 
\begin{equation}
A_{K}\equiv N \frac{\sigma _{K}^{2}}{<K>^{2}}
\propto \frac{\sigma _{K}^{2}}{T}
\label{eqAk}
\end{equation}
In this way, the system specific heat can be calculated through Eq.~\ref
{eqFluc} knowing both, $<K>$, and $\sigma _{K}^2$. 

Since the dynamics of the expansion 
tends to drive the system into a rather rarefied state, and 
allows event by event volume 
fluctuations which are suppressed by the sharp boundary, 
it is interesting to see whether
such abnormal fluctuations can be reached in a fully dynamical simulation.
In this case the average kinetic energy cannot be associated to a temperature
any more, and the relationship expressed in
Eq.~\ref{eqFluc} can no longer be invoked to estimate the system specific
heat. However meaningful information can still be obtained from 
$A_{K}$ when we resort to dynamical quantities only. In Figure~\ref{figakt}
we show the time evolution of MST mass spectra (panels \ref{figakt}.a, \ref
{figakt}.b, and \ref{figakt}.c), $A_{K}$ (panels \ref{figakt}.d, \ref{figakt}%
.e, and \ref{figakt}.f), $K$, and $\sigma _{K}$ (panels \ref{figakt}.g, \ref
{figakt}.h, and \ref{figakt}.i) calculated for a system initially confined
at a density $\rho =\rho _{3}=0.6\sigma ^{-3}$, for three different
energies. We also show in panels \ref{figakt}.d, \ref{figakt}.e, and \ref
{figakt}.f, the corresponding time evolution of the normalized mass of the
biggest MST fragment (dashed line).

From this figure we can see that the time at which $A_{K}$ attains its
asymptotic 
value ( panels d, e, and f) signals the time at which, in average, the fragmentation 
pattern is settled for each event of a given energy and initial density values. 
To support this idea, in the upper panel of Figure~\ref{figakt} 
we show MST spectra calculated at different times. It can be
noticed that at $t\sim 40 t_0$ a significative proportion of the mid-size 
asymptotic MST-fragments are already produced, i.e. internal surfaces already appeared.
For later times the big residue evolves just by light particles evaporation.

As a result of the expansion, physical observables like $A_{K}$ change in
time during a relatively short lapse after which they attain almost constant
values. A {\em freeze-out} time, $\tau_{fo}$, can then be introduced, after of 
which the evolution becomes almost irrelevant. 
Moreover, Figures~\ref{figakt}.d,~\ref{figakt}.e, and \ref{figakt}.f 
also indicate that asymptotic measurements
allow to reconstruct quantities as early as $\tau _{fo}\sim 40t_{0}$.

This observation agrees with previous results. In reference~\cite{ale99} it was
shown that for unconstrained systems a time of fragment formation ($\tau
_{ff}$), related to the stabilization of microscopic clusters composition,
 can be defined. 
 For the cases studied we found that $\tau _{ff} \sim \tau_{fo}$,
 meaning that the information is frozen once the fragment surfaces 
have developed.
From the lower panels we see that the variance of the
kinetic energy,  $\sigma _{K}$, also attains its asymptotic value at this time.
It is important to stress that this definition of a freeze out time
corresponds to a MaxEnt principle within the constraints imposed
by the dynamical evolution. As such, this definition
does not imply that the freeze out configuration corresponds to an absolute
entropy maximum, i.e. to a thermodynamic equilibrium.

For the lowest considered energy, $\sigma _{K}$ 
does not change significantly in time
and this stays true if we change the initial density. This
is the expected behavior as the system, in almost every realization,
only evaporates light clusters in its early evolution, and a big fragment 
is always present almost independent of the system volume. 
On the other hand, for the more energetic cases the initial decrease 
of $\sigma _{K}$ can be related to the preequilibrium dynamics that 
involves a flux development (expansion) and prompt emission of light particles.
Afterwards, the development of internal surfaces gives rise to a large set
of possible decaying channels (opening of phase space) that is responsible 
for the $\sigma _{K}$ enhancement.

$A_{K}$ asymptotic values characterize the number of 
different fragmentation configurations compatible with the total
energy constraint.  Large kinetic energy fluctuations
reflect large potential energy fluctuations that can be directly related to
event by event final cluster distributions. In this sense $K$ fluctuations
can be associated with the available number of decaying channels for a given
energy, i.e. with the microcanonical entropy. 

\begin{figure}[tbp]
\epsfxsize=6cm \center{\leavevmode\epsfbox{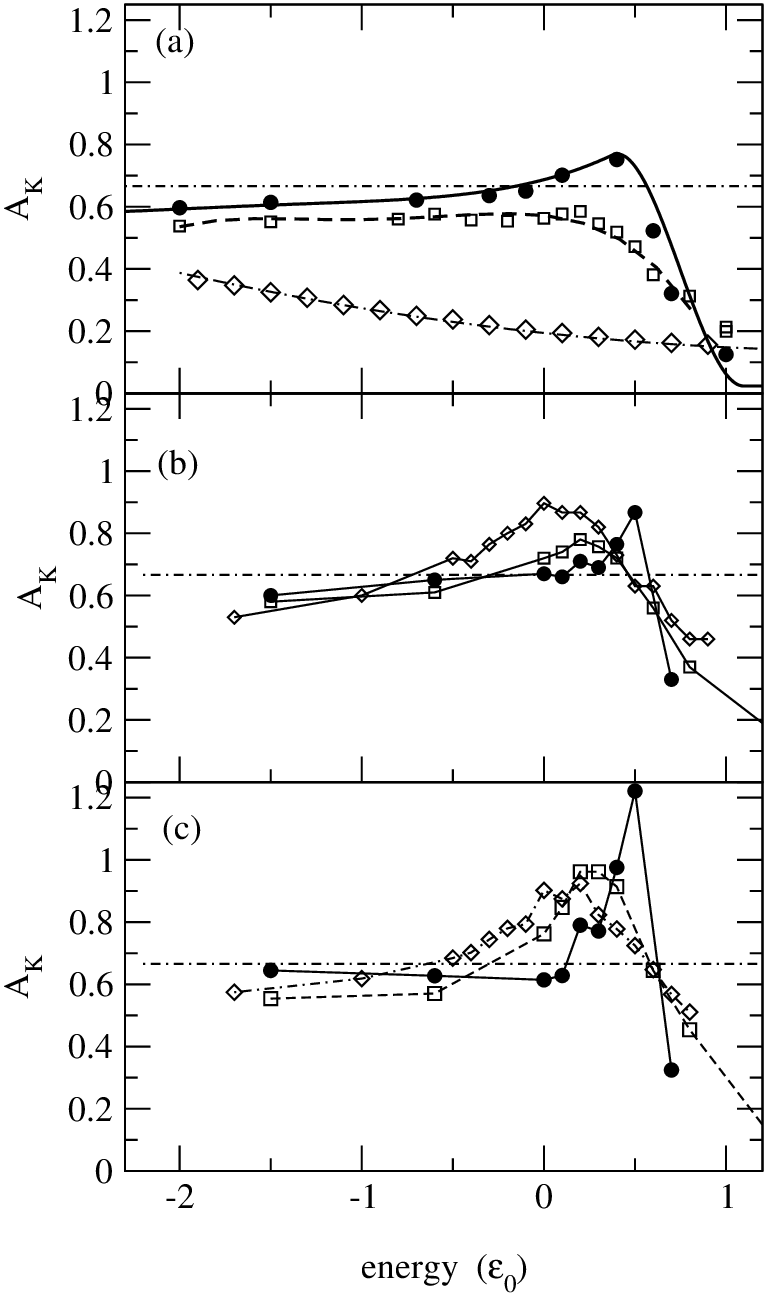}}
\caption{Top: Relative kinetic energy fluctuations as a function of the total
 energy for densities $\rho_1$ (circles), $\rho_2$ (squares), and $\rho_3$ 
(diamonds). 
Middle: same as top, but evaluated at the time of $\sigma_K$ 
stabilization $(t(\rho_1)\sim 5 t_{0}$, $t(\rho_2)\sim 30 t_{0}$, 
$t(\rho_3)\sim 40 t_{0})$. Bottom: same as top, but evaluated
for asymptotic configurations.
See text for details.}
\label{figEnh}
\end{figure}

\section{Discussion}
We summarize in Fig.\ref{figEnh} the results of $A_K$ calculations obtained 
so far.  In the upper panel we show the fluctuations in kinetic energy for the 
three confined states analyzed at densities $\rho_1$ (circles), $\rho_2$ 
(squares), and $\rho_3$ (diamonds). The solid lines are not a fit, 
but an $A_K$ estimation using information from the respective system's caloric 
curve.  
The nice agreement between the lines and the symbols demonstrates
the quality of our numerical sampling. 
In the middle and lower panels we report the same quantity calculated 
at the time of $\sigma_K$ stabilization (as was mentioned above, this time
is well-correlated with the time of fragment formation) and for the 
corresponding asymptotic configurations respectively. 
In the upper panel the horizontal 
dotted-dashed line depicts the expected canonical value for $A_K$. 
We included it also in the middle and bottom panels, only as a reference value.
In the fluctuation analysis performed on experimental heavy ion 
data, the average kinetic energy is approximately corrected for evaporation 
and for the Coulomb boost, while fluctuations are assumed to be correctly
given by the asymptotic measurement~\cite{cvdagostino}. 
This means that the middle
panel is the most relevant for a comparison with experimental data.

The confined state with density $\rho_1$, that belongs to region A in 
the phase diagram of Fig.\ref{figES} (full circles) displays 
a maximum that is well above the canonical value.  
As such there is a region for which the thermal response
function is negative. On the other hand for the
confined states $\rho_2$ and $\rho_3$, that belong to region B and C 
respectively of the phase diagram, the
corresponding values of the kinetic energy fluctuations are always below the
canonical value. 

If we now turn our attention to the results of the same calculation but now
performed on the states resulting from removing the confining walls and
allowing the system to expand and fragment, the results displayed in the
middle and lower panel of Fig.\ref{figEnh} are obtained.
Almost independent of the initial condition, the normalized fluctuations
fall on a curve which is mainly determined by the total energy and fairly
close to the thermal response of an initially diluted system in microcanonical
equilibrium.
The kinetic energy fluctuations
that were below the threshold for all energies have been clearly enhanced
in the energy region defined by $0\leq E\leq 0.5 \epsilon_0$. 
This feature, as was already mentioned, reflects the existence of
quite different available decaying channels that are explored by the 
systems dynamics. In particular, at these energies,
partitions with big residues on one hand, and high multiplicity partitions
with no such big structures on the other, can be produced giving rise to 
power-law like distributions.
 If we look at the asymptotic stage (lower panel), the picture is somewhat
blurred by secondary evaporation that continuously decreases the average kinetic
energy without affecting th fluctuation. 
It is worth noticing that at time of fragment formation the average system
volume is
approximately independent of the initial density: this means that for all
initial conditions the information is frozen when the system reaches the region
of the phase diagram (region $A$ of figure ~\ref{figES}) where abnormal
fluctuations are expected in equilibrium, i.e. in the case of a full opening 
of the accessible phase space. This observation may suggest that the 
collective motion leads the system to a low density freeze out configuration
which is not far away from a thermodynamic equilibrium. However, to make any
conclusion about the possible degree of equilibration and the characteristics of
the relevant statistical ensemble, a more detailed analysis is needed. 

\section{Conclusion}
In conclusion, in this work we have shown that for free to expand fragmenting
systems, initially equilibrated within sharp boundary conditions,
an intermediate stage exists
where new microscopic correlations arise. This happens in the presence of a 
collective expansive motion that induce internal surfaces to appear in the 
system, preceding the formation of well defined asymptotic clusters.

In addition, the analysis of the dynamical evolution of $A_K$ turned out
to be fruitful. On one hand it allowed us to give a physically 
meaningful estimation of the freeze-out time, which turned out to be in good 
agreement with the corresponding fragment formation time. 
On the other, it was shown that a link can be established between
asymptotic values of that observable and mass distribution functions. In isolated 
systems, like the one studied here, a maximum of $A_K$ as a function of the system
energy can be related to a noticeably enlargement of the number of possible decaying
 channels.
This enhancement of the relative kinetic energy fluctuation can thus be associated
with an opening of phase space that is expected in phase transitions taking place in 
finite isolated systems.

It is worth noting that even if the non-constrained fragment formation stage
seems to provide a quite effective memory loosing mechanism (see Fig.
\ref{figecramst}), preliminary results support the idea that information on
the initial system density can still be retrieved from certain asymptotic cluster
configurational features. This analysis, along with a thorough characterization of
the system equilibration properties at $\tau_{fo}$ is currently under progress.

This work was done with partial financial supports from UBA and CONICET. 

\bibliographystyle{ieeetr}
\bibliography{paperAk}

\end{document}